# Demonstration that the Force vs. Extension Curves for Overstretched Double Stranded DNA Depend on which ends are Pulled

C. Limouse, C. Danilowicz, V. W. Coljee, N. Kleckner[1], M. Prentiss*

Department of Physics, Harvard University, Cambridge, MA 02138, USA.

[1] Department of Molecular and Cellular Biology, Harvard University, Cambridge, MA 02138, USA.

Contact Details:
Tel: 1-617-495-4483
Fax: 1-617-495-0416
E-mail: prentiss@fas.harvard.edu

PACS Codes:  87.14.Gg, 87.15.-v, 82.35.Rs, 82.37.Rs

## Abstract

We present the first comparison of the force vs. extension curves that result when dsDNA is overstretched from the 5'5', 3'3', and 3'5' ends. In high salt buffers, the curves for all pulling techniques are similar and show no hysteresis. In low salt buffers, the curves for all pulling techniques are similar and show a marked hysteresis; however, in 150 mM NaCl, there are strong differences between the decreasing force curves for different pulling techniques. We confirm that the overstretched state is not single stranded and note that if the hysteresis is due to inadequate charge screening of the repulsion between phosphates on opposite strands, these results are consistent with existing theory that predicted that the overstretched states are forms of double stranded DNA that depend on the pulling technique, where 5'5' stretching produces a narrow fiber and 3'3' stretching produces a more widely spaced ladder.

Double stranded DNA (dsDNA) exists in a variety of different conformations in vivo, even in the absence of force. It is well known that applied force and torque in vivo and in vitro can lead to changes in the conformation of dsDNA, some of which have been explored using single molecule techniques that stretch and/or twist single dsDNA molecules [1-10]. Recent work has shown that stretching dsDNA exerts a torque that will induce the strands to rotate if there is a nick in the backbone [11]. It has long been known that if a larger force is applied to the ends of dsDNA, the molecule will undergo an overstretching transition into a structure having a length per base pair that is approximately 1.7 times the length of the canonical B-DNA [1-10]. The RecA protein is believed to produce a similar reaction in vivo playing an important step in the repair of damaged DNA [12]. In this paper, we present a comparison between the conformation of dsDNA that is created by applying a force of approximately 65-70 pN to the 5'5' ends of dsDNA molecules and the conformation that is created when the same sequence is stretched from the 3'3' ends and demonstrate that the structure of the overstretched state does indeed depend on the ends to which the force is applied.

Earlier theoretical work had proposed that overstretched DNA is fully double stranded DNA. In the overstretched structures, the spacing between phosphates on a given backbone was conserved; however, the change in base pair tilt and helical pitch results in a large change in the length per base pair [13]. That work suggested that the structure of the overstretched state obtained by pulling the DNA from the 5'5' ends is different than the state obtained by pulling the DNA from the 3'3' ends. This difference is a consequence of the base pair tilt in B-DNA. When DNA is overstretched by pulling from the 5'5' ends, the base pair tilt is increased whereas when the DNA is pulled from the 3'3' ends the base pair tilt is decreased and then reversed. In addition, when the DNA is stretched from the 5'5', the spacing between the two phosphate backbones is much smaller than the spacing between the backbones in B-DNA and the structure resembles a tightly twisted fiber. In contrast, when dsDNA is stretched from the 3'3' ends, the overstretched structure resembles a ribbon with almost no twist, where minimum spacing between phosphates on opposite backbones is larger than for the structure obtained when the same molecule is pulled from the 5'5' ends. These theoretical predictions [13] had not been tested until this work.

Previous experimental work has shown that there are conditions under which the overstretched structures do not immediately relax back to the B-form when the applied force is decreased below the overstretching transition. [2,6,14,15]. Thus, the force vs. extension curve for overstretched DNA displays hysteresis. It had been suggested that this hysteresis was the result of melted DNA that does not reform into B-DNA [9] possibly because of the presence of nicks in the structure [16,17]. Later in this paper we will show that if the maximum applied force is less than 1.2 times the overstretching force, the ssDNA fraction above the overstretched state is too small to account for the observed hysteresis; therefore, the hysteresis is not due to the formation of melted DNA. Instead, the hysteresis is a consequence of the metastability of the double stranded conformation created by overstretching the DNA.

Recent experimental work has used measurements of the hysteresis that results when dsDNA is overstretched by pulling on the 3'3' ends to determine the screening effectiveness of different cations [18]. In that work, the hysteresis as a function of ion concentration was measured for several different cations. Ions with the same melting temperature did not necessarily have the same hysteresis; however, the hysteresis for all monovalent ions decreased with increasing ion concentration. These two results suggest that the hysteresis measures the effectiveness of ions in screening the charges in the conformation of dsDNA that results when the molecule is overstretched by pulling from the 3'3' ends.

In this work, we consider one single salt, NaCl, at several concentrations and measure the screening effectiveness of this salt for the structures that result when DNA is overstretched by applying a force to the 3'3', 5'5', and 3'5' ends. If all of the structures resulting from overstretching were the same, screening should also be the same; however, we will show that for 150 mM NaCl the screening is not the same: the hysteresis for structures pulled from the 3'3' ends are consistently smaller than those obtained when the dsDNA is pulled from the 5'5' ends. These differences between pulling techniques imply that the DNA is bound by single binding sites at the end, rather than by multiple sites as in previous work [1, 9, 10] or random non-specific binding to the surface. In addition, the consistency of the results shows that random nicks do not play a significant role since the nicks would redistribute the force randomly across the backbones, resulting in the same hysteresis regardless of the end to which the force was applied. Finally, if, as the data from the cation experiments suggests [18], the hysteresis is associated with a failure of screening at a length scale shorter than the minimum distance between phosphates on opposite strands, then our results imply that there is a characteristic spacing in DNA overstretched from the 5'5' ends that is smaller than the characteristic spacing for DNA overstretched from the 3'3' ends. This result is consistent with the overstretched structures proposed by Lavery and Lebrun [13]

**Materials and Methods**

A magnetic tweezers set up was used to stretch lambda phage dsDNA attached by one single bond at each end in one of three configurations: 3'3', 3'5', and 5'5'. In each configuration, one single biotin label was present at one end and one single digoxigenin label at the other end of lambda phage dsDNA. In order to be able to apply forces well above the overstretching force, lambda DNA was modified following two additional strategies: a) a biotinylated oligonucleotide was hybridized and annealed to one end whereas the opposite end was modified by incubating the construct with biotin-11-dCTP, dATP, dGTP, dTTP, and Klenow exo- polymerase for one hour at 37ºC yielding a 3'3' construct presenting one and three biotin labels at each end, respectively; b) a 3'5'construct was prepared by hybridizing and annealing oligonucleotides containing three biotin labels each.

The experimental procedure was described in detail in previous work [19]. Dynal beads of 4.5 μm in size were prepared by reacting tosylactivated beads with streptavidin; forces as high as 250 pN can be applied using these beads in our current apparatus.

**Results**

A. Overstretching by pulling from different ends.

In the first section of this paper, we will confine ourselves to the case where the maximum applied force is less than 1.2 times the overstretching force, $F_{ov}$, which will turn out to be the force regime where the fraction of ssDNA formed by overstretching is very small. Figure 1 shows typical force vs. extension curves for nine individual molecules. The curves were obtained by applying a constant force for 1 second, and then increasing or decreasing the force by a fixed step of 0.2 pN at lower forces, or 0.05 pN around the transition, after which the force was then left constant for another dwell time of 1 sec. For the data shown in Fig. 1, all of the extensions were recorded at the end of the dwell time. Each column from left to right includes results for each pulling technique: 3'3', 3'5', and 5'5', respectively, whereas each row (top to bottom) corresponds to a specific salt concentration: 1M, 150 mM, and 20 mM NaCl in 10 mM phosphate buffer pH 7.4, respectively. In all cases, at low forces, the decreasing force curve overlaps the increasing force curve for the first cycle which also overlaps the increasing force curve for the second cycle, suggesting that almost the entire molecule has returned to its original state when the force is decreased.

Each curve represents the results for a different single dsDNA molecule attached by a different magnetic bead, but the results are typical. For a given buffer, the variation in $F_{ov}$ for the different pulling techniques was less than 1.2 % of the $F_{ov}$. This variation is comparable to the variation within a given pulling technique. Similarly, the variation in extension between pulling techniques was less than 3% of the extension, which was also comparable to the variation within a given pulling technique. The first row in Fig. 1 shows the force vs. extension curves for molecules in 1 M NaCl. For a 1 M NaCl solution all of the overstretching transitions are reversible and there are no metastable intermediate states. The third row in Fig. 1 shows the force vs. extension curves for molecules in 20 mM NaCl where all three pulling techniques show hysteresis. In contrast, in 150 mM NaCl, the observed hysteresis depends strongly on the pulling techniques.

We quantify the hysteresis by defining $H_{Low}$ such that:

$$H_{Low} = \int_{F_{min}}^{F_{max} < 1.2 F_{ov}} x(F) \, dF \qquad (1)$$

where x(F) is the extension as a function of the applied force F, and $F_{ov}$ is the overstretching force. A summary of the resulting hysteresis values is shown in Table 1. The differences in $H_{Low}$ for the 3'3' and 5'5' curves (0.3 ± 0.8 and 13 ± 8 pN μm) exceed the standard deviation in each group. The 3'5' construct yields an intermediate area difference of 6 ± 4 pN μm

The variation in the curves as a function of the ionic strength for the different pulling scenarios is in good agreement with the naïve hypothesis that a metastable state exists if the Debye length is longer than the minimum spacing between phosphates on opposite backbones given that the Debye lengths in 20 mM, 150 mM, and 1 M NaCl are 2.1 nm, 0.77 nm, and 0.3 nm, respectively, and the minimum spacing between phosphates on opposite backbones is approximately 0.35 nm for 5'5' pulling and 0.8 nm for 3'3' pulling [13,20].

Previous theoretical work predicted that the overstretched structure would depend on sequence as well as pulling technique [13,20]. This effect would be most evident for regular periodic sequences as has been further shown in AFM experiments [14]; however, it is possible that overall GC content could have an effect. In order to study whether there were any effects associated with overall GC content, we measured the hysteresis for two fragments of lambda phage DNA corresponding to the GC-rich region (10 kb) at one end of the molecule and the AT-rich region (14 kb) at the other end. These fragments were pulled from 3'5' ends. The resulting force vs. extension curves differ significantly from those for full lambda: the GC-rich fragment displays no hysteresis and the AT-rich fragment showed a large hysteresis comparable with the total hysteresis observed when pulling 3'5' full lambda dsDNA. If the hysteresis for a long fragment is simply the linear sum of the hysteresis for its components, then this result would imply that the 14 kb AT rich region accounts for most of the hysteresis observed when the complete lambda phage molecule is overstretched.

B. Overstretching and applying forces above $2F_{ov}$

In order to distinguish the hysteresis considered so far in this paper from the hysteresis associated with creating ssDNA, we have deliberately generated ssDNA by applying maximum forces larger than those used in the previous section. Earlier work has conclusively shown that when DNA is pulled from the 3'5' ends, ssDNA can be generated if the applied force exceeds the overstretching force by a factor of approximately 2 [14,21]. The ssDNA is generated because the 3'5' labels attach one ssDNA strand between the force transducer and the surface, whereas the other ssDNA strand that makes up the original dsDNA is not attached to any surface; therefore, when the shear force is sufficiently high, the two ssDNA strands separate and the unattached strand diffuses away leaving a single stranded DNA molecule bound between both surfaces. If the molecule only partially converts to ssDNA and does not reanneal when the force is reduced below the overstretching transition, then the formation of ssDNA will indeed produce a hysteresis in the force vs. extension curves.

In order to distinguish the hysteresis due to the formation of ssDNA from $H_{Low}$, the hysteresis studied in the previous section, we define a new quantity $H_{High}$ as follows:

$$H_{High} = \int_{1.2F_{ov}}^{F_{max}} x(F)\, dF \qquad (2)$$

where H $_{High}$ measures the hysteresis in the high force region only, whereas H$_{Low}$ considered above corresponds to the hysteresis for the case where F$_{max}$ < 1.2 F$_{ov}$.

Figure 2(a) shows the force vs. extension curves for DNA pulled from the 3'5' ends through several cycles. In the figure, each overstretching cycle is distinguished by color with the solid line showing the increasing force curve and the dotted line showing the decreasing force curve. The first cycle is shown by the red curve; the maximum force was 127 pN, and H$_{High}$ is 0, though H$_{Low}$ is 37 pN µm. It is well known that the worm like chain is a good model for the force vs. extension curve for dsDNA below the overstretching transition [22]:

$$WLC(F) = b_{ds\,max}\left(1 - \frac{1}{2}\left(\frac{k_b T}{F\,P_{ds}}\right)^{1/2} + \frac{F}{S_{ds}}\right) + ds_{offset} \qquad (3)$$

where b$_{ds\,max}$ is the dsDNA contour length, $k_B$ is Boltzmann's constant, P$_{ds}$ is the persistence length, S$_{ds}$ is the elastic stretch modulus, and ds$_{offset}$ is a constant offset that can arise if the DNA is not bound exactly to the edge of the capillary that is imaged by the camera. A fit to the ascending force curve for the first cycle shown in Fig. 2(a) gives b$_{ds\,max}$ = 15.1 µm, and S$_{ds}$ = 700 pN. In the absence of data at very low forces, the fits were insensitive to the value of P$_{ds}$, so we used P$_{ds}$=50 nm, which is consistent with the results in Ref. [22]. The second cycle is shown by the green curve during which the maximum force is increased to 148 pN; there is a clear partial transition to ssDNA and H$_{High}$ is 30.4 pN µm. Note the change in slope for the decreasing force curve in the region above the transition, where the reduced slope is characteristic of ssDNA and the increasing force curve clearly follows the higher slope characteristic of dsDNA. For the second cycle, there is a clear difference in the force vs.

extension curve for the increasing and decreasing force curves even in the force range between 70 pN and 127 pN, where the first cycle shows no difference. Fits to the first cycle suggest that the ssDNA formed by applying a maximum force of 127 pN is negligible, whereas one can see from the figure in the region above the overstretching transition the force vs. extension curve for the second cycle is half way between the ssDNA curve and the dsDNA curve, suggesting that 50% of the molecule has made the transition to ssDNA.

The blue curve shows the third cycle. This cycle has a maximum applied force of 165 pN and exhibits a complete transition to ssDNA allowing the unlabeled strand to diffuse away. The transition is not quite complete when the force begins to decrease, but just after the force decrease begins the molecule completes the transition to one single ssDNA molecule; consequently, the increasing force curve for the fourth cycle (purple) clearly follows the curve for the third cycle (decreasing force) confirming permanent ssDNA formation.

For the fourth cycle $H_{Low}$ and $H_{High}$ are both zero since the force vs. extension curve simply corresponds to the force vs. extension curve for ssDNA. It is well known that the force vs. extension curve for ssDNA is well described by a modified freely jointed chain [23] as follows:

$$dx(F) = -L\left(-\coth\left[\frac{bF}{k_BT}\right] + \frac{k_BT}{bF}\right) * \left(1 + \frac{F}{S_{ss}}\right) * 48500 + S_{ss\,offset} \qquad (4)$$

where $k_B$ is Boltzmann's constant, b is the Kuhn length, L is the distance between base pairs, and $S_{ss}$ is the modification of the freely jointed chain that allows for additional increases in length at high forces due to bond elasticity. A fit to the fourth force vs. extension cycle gives L = 0.55 nm, b = 1.18 nm, Sss = 690 pN. The differences between the force vs. extension curves for ssDNA and the force vs. extension curves for molecules overstretched to maximum forces less than 1.2 $F_{ov}$ are quite clear from Fig. 2(a).

Given the force vs. extension curves for dsDNA and ssDNA, it is possible to fit the shear transition that occurs during the increasing force portion of cycle 3 as a linear combination of ssDNA and dsDNA. Figure 2(b) shows a fit to the shear transition from dsDNA to ssDNA that is shown in Fig. 2(a), where the fraction of ssDNA as a function of force is given by Eq.5:

$$ssDNA_{fraction}[F] = frac_{inf\ inity} + \left[\frac{frac_0 - frac_{inf\ inity}}{1 + \exp\frac{-F_c + F}{\sigma}}\right] \quad (5)$$

where $frac_{infinity}$ is the fraction of the molecule in the ssDNA state when the force approaches infinity, $frac_0$ is the fraction of the molecule in the ssDNA state at forces below the shear transition but above the overstretching transition. We assumed $frac_{infinity}$ =1 and $frac_0$ =0. The data shown in Fig. 2(b) was fit using Eq.5 and yielded $F_c$ =152 pN and $\sigma$ = 12 pN. Notice that the fit indicates that the fraction of ssDNA at forces below 1.2 $F_{ov}$= 78 pN is less than 0.003 and the averaged measured value is 0.02 ± 0.02.

Figure 2(c) shows the data for the third cycle (blue circles on solid and dotted lines) that is shown in Fig. 2(a), along with the predicted force vs. extension curves based on the fits to the WLC and mFJC models. In Fig. 2(c), the solid circles show the measured values for the increasing force curve, and the open circles show the corresponding values for the decreasing force curve. The light blue dotted lines in the figure show the worm like chain fit to the increasing force curve of the first cycle below the overstretching transition and the same curve displaced to match the overstretched length. For forces below the overstretching transition, the fit to the WLC is good. For forces between 70 and 130 pN, there is good agreement between the displaced WLC and the data.

In contrast, the force vs. extension curve for ssDNA that is shown by the red dotted line does not fit the force vs. extension curve for the increasing force curve just above the overstretching transition, as can be seen very clearly in the figure; however, the ssDNA curve is a good fit to the decreasing force curve above the overstretching force. The data presented in Figs. 2 (b) and (c) provide additional proof that at forces just above the overstretching transition the structure is dominantly double stranded DNA. The overstretched DNA clearly does not become dominantly single stranded until the shear force exceeds twice the overstretching force, $F_{ov}$. This point was made in earlier theoretical work [16] where it was noted that the measured slope of the force vs. extension curve just above the overstretching transition does not match the slope for ssDNA. The measured force vs. extension curves in Refs. [8] and [21] also clearly show that the slope of the force vs. extension curves at forces just above the overstretching transition is much steeper than the slope for ssDNA. In addition, recent experimental work explicitly distinguished the overstretching transition, which occurs at approximately 65 pN, from the shearing transition that occurs at a much higher force [8].

It has been suggested that even if the overstretched state does not consist of ssDNA the presence of nicks in the backbone can produce a hysteresis in the force vs. extension curves because ssDNA formed near the nicks does not reannneal [16,17]. If the hysteresis were due to ssDNA that does not reanneal, then the fraction of ssDNA that occurs above the overstretching transition must be at least as large as the ratio of the molecule that displays hysteresis. Thus, if as in the first cycle shown in Fig. 2(a), 13% of the molecule does not return to the B form, this would require that at least 13% of the overstretched molecule be ssDNA since the amount of ssDNA that fails to reanneal cannot exceed the fraction of ssDNA present in the overstretched molecule, whether nicks are present or not. A fit of a linear combination of ssDNA and dsDNA to this force vs. extension curve gives an ssDNA fraction of 2 ± 2 %; therefore, the observed hysteresis must not be due to the presence of ssDNA that does not reanneal. Additional measurements on more than 100 molecules have consistently shown that for all pulling techniques the measured fraction of ssDNA at 80 pN is always less than 5%; therefore, the result for Fig. 2(b) can be generalized: the observed hysteresis must not be due to the presence of ssDNA that does not reanneal, whether there are nicks present or not. Even when ssDNA is present, as in the second cycle shown in Fig. 2(a) where the ssDNA fraction is 50%, the 65% of the DNA that does not return to the B form when the force is reduced to 60 pN is larger than the 50% of the DNA that made the transition to ssDNA, demonstrating that even for a maximum applied force greater than 1.2 $F_{ov}$, the ssDNA fraction is not sufficient to account for the observed hysteresis.

If one is interested in additional experimental studies of the role played by ssDNA in $H_{Low}$, it would be useful to have an experimental probe that can determine the fraction of ssDNA that is present when the maximum applied force is less than 1.2 $F_{ov}$ using a technique other than curve fitting to a linear combination of ssDNA and dsDNA since such fits are not sensitive to very small fractions of ssDNA. If results using this probe are to be meaningful, it is necessary to show that the probe can indeed sense the presence of ssDNA created by applying a shear force to the DNA. A useful test would be to create ssDNA using shear force, as shown in Fig. 2(a), and then demonstrate that the ssDNA thus created can be measured by the probe. Earlier work by the Bensimon group used the hysteresis in the force vs. extension curves induced by the presence of glyoxal to demonstrate that ssDNA was created by the mechanical manipulation of the molecule [22] where the hysteresis was the result of the glyoxal binding to open ssDNA created by torque. Glyoxal was also used to block hairpin formation in ssDNA [23, 24]. In the Bensimon work [22], the glyoxal interacted with the dsDNA distorted by torque for one hour; in this work a probe that binds to ssDNA created by applying a shear force within less than a minute is required. It was therefore crucial to prove that glyoxal can immediately bind to ssDNA that is created by a shear force if we are to use glyoxal to determine whether ssDNA is created by overstretching. Figure 2(a) showed that ssDNA can be created by applying a shear force to the 3'5' ends even in the absence of glyoxal, so this experiment does not allow for conclusive demonstration of glyoxal reactivity. In order to prove that glyoxal can maintain ssDNA created by shear stress, one needs to perform an experiment where no permanent ssDNA is created in the absence of glyoxal. Thus, one would like to repeat the experiment done in Fig. 2(a) in a system where the second ssDNA strand cannot diffuse away so that in the absence of glyoxal the molecule always returns to dsDNA at sufficiently low forces. The force vs. extension curves for such a system, are shown in Fig. 3(a). Previous work had shown that a Klenow enzyme can create dsDNA with multiple biotin labels attached to one strand [1]. We have constructed DNA molecules where one end was bound to a capillary surface by multiple biotin labels attached to dsDNA using the Klenow polymerase, and the other 3' end was attached to a short ssDNA complementary sequence containing one biotin label that was in turn bound to a magnetic bead. When a shear force above 2 $F_{ov}$ is applied to this molecule, it will separate into two strands of ssDNA except at the end bound to the capillary surface, where both strands remain attached to the capillary surface by the dsDNA Klenow handle. A molecule attached using this technique can thus complete the shear transition at forces above $2F_{ov}$, but return to dsDNA at low forces because both strands remain bound together at the capillary surface. This effect is illustrated in Fig. 3(a), where successive force vs. extension cycles are shown. An increasing ssDNA fraction is created as higher shear forces are applied, resulting in significant $H_{High}$; however, when the force is decreased below 20 pN, dsDNA reforms and the increasing force curves always overlap the dsDNA curve. The first force vs. extension cycle is shown in red, where the solid line shows the curve for increasing force and the dotted line shows the curve for decreasing force. The molecule clearly begins as dsDNA and undergoes the overstretching transition at approximately 65 pN. The transition to ssDNA begins at about 130 pN. The force is increased up to 147 pN, and decreased again. This force is just enough to make approximately 25% of the molecule into ssDNA. There is a clear hysteresis in the high force region associated with this ssDNA formation where the decreasing force curve above the transition has a lower slope than the increasing force curve; however, when the force is reduced to 20 pN, the molecule has returned to dsDNA. The second overstretching cycle is shown in green. The increasing force curve for the second cycle clearly follows the increasing force curve for the first cycle, indicating that the DNA has returned to its original state (dsDNA). In the second cycle the maximum force is 165 pN, which is more than enough to transform the molecule to ssDNA; however, the Klenow filled end keeps the two strands together, and at 20 pN, dsDNA is again recovered. The increasing force curve for the third cycle shown in blue almost perfectly overlaps the increasing force curves for the second and third cycles showing that even when the rest of the molecule had completed the shear transition dsDNA is reformed at low force due to the dsDNA present at the capillary surface. This complete reannealing of dsDNA at low forces occurs in all of the molecules that we examined for the same construct.

If the experiment shown in Fig. 3(a) were repeated in the presence of glyoxal, one would expect that the molecule would not return to dsDNA at low forces if the glyoxal were able to react immediately and irreversibly with ssDNA. Fig. 3(b) shows the force vs. extension curves for a molecule bound with the same construction used in Fig. 3(a), but now the experiment is done in the presence of 0.5 M glyoxal. Several force vs. extension cycles for the same molecule are shown; however, in contrast with Fig. 3(a) where the molecule always returned to dsDNA at low forces and the increasing force curves were identical for all cycles, in the presence of glyoxal, ssDNA persists even at the lowest forces. The increasing force curves differ as increasing fractions of the molecule are converted to ssDNA by increasing the final force. Again, the first force vs. extension curve is shown in red with the increasing force curve shown by the solid line and the decreasing force curve by the dotted line, where the low force curves for increasing and decreasing force overlap. If ssDNA had been created by the interaction with the glyoxal before the force was applied, then this first curve would show an ssDNA fraction, but it does not. Notice also that the overstretching force for the first force vs extension curve is not shifted by the presence of glyoxal, suggesting that it has not affected the stability of B-DNA. The second, third, and fourth cycles are not shown for clarity but are used to calculate the ssDNA fraction shown in Fig. 3 (c). The green line in Fig. 3(b) shows the fifth cycle where the maximum applied force of 130 pN is sufficient to create about 15% ssDNA, which remains even when the force is decreased to 20 pN. This clearly indicates that within the time of our experiment glyoxal can bind irreversibly to ssDNA created by applying a shear force and maintain the open ssDNA even when the force is reduced to 20 pN. The seventh cycle is shown in blue. Above the overstretching transition, most of the DNA is in the ssDNA state. For this cycle, slightly more than half of the DNA is maintained in the ssDNA state by the glyoxal even when the force is reduced to 20 pN again confirming that glyoxal can bind to DNA opened by a shear force and maintain the open DNA even when the applied force is smaller than the force at which dsDNA would reform in the absence of glyoxal. Experiments on other molecules have shown that it is the maximum applied force, not the number of force vs. extension cycles that determines the ssDNA fraction that is held open by glyoxal.

One can again fit the force vs. extension curve to a linear combination of the force vs. extension curve for ssDNA and dsDNA as a function of increasing force. The results are shown in Fig. 3(c), where unlike the case in the absence of glyoxal, the fraction of ssDNA does not approach infinity as the force approaches infinity. A fit to the phase transition equation (Eq. 5) represented by the dotted line of the figure gives $F_c$= 130.7 pN, $\sigma$ =11 pN, and frac$_{infinity}$ = 0.53. This data provides additional evidence that the overstretching transition does not create significant regions of ssDNA and that the overstretched state is an alternate form of dsDNA, not melted DNA.

C. Glyoxal used as a probe for the overstretched structures obtained by pulling from different ends.

Having established that glyoxal can bind to ssDNA created by a shear force within the time of our forward and reverse scans, we will use glyoxal to probe the ssDNA fraction that is induced by applying a shear force to the 3'3', 3'5', and 5'5' ends in a buffer that contains glyoxal. We will show that overstretching from the 5'5' ends to a force below 1.2 $F_{ov}$ creates a small fraction of ssDNA to which the glyoxal can bind, and that this binding reduces the hysteresis. We observe no effect on the overstretching curves for 3'3' and 3'5' when the shear force is less than 1.2 $F_{ov}$. Figure 4 shows the force vs. extension curves in 150 mM NaCl in the presence of two concentrations of glyoxal. A typical force vs. extension curve for a DNA molecule in 10 mM glyoxal, when the molecule was pulled from the 5'5' ends to a force just above the critical force for the overstretching transition, is shown by the red curve and compared with another molecule pulled from the same ends in the absence of glyoxal (green curve). The red curve shows almost no hysteresis and is characteristic of the results for glyoxal concentrations between 10 and 50 mM (blue curve) that show an average $H_{Low}$ = 1.4 ± 2.4 pN μm; at concentrations below 10 mM the hysteresis is reduced but not eliminated. The fact that the red and blue curves differ from the green curve indicates that the glyoxal was able to bind and did have an effect on the timescale of the experiment. This overstretching of dsDNA using the 5'5' ends results in a small fraction of ssDNA that is available to bind with glyoxal resulting in a reduction in the hysteresis. In contrast, when dsDNA is pulled from the 3'5' ends in 10 mM glyoxal (see below), the hysteresis is unchanged and similar to the curve shown in Fig. 1. This result suggests that in the case where dsDNA is pulled from the 3'5' ends either no ssDNA is available to bind with glyoxal or that the binding of glyoxal does not affect the hysteresis. This is in marked contrast with the result obtained when pulling from the 5'5' ends where curves in the absence of glyoxal were quite hysteretic, but the curves in 10 mM glyoxal showed no hysteresis offering further proof that there are structural differences in the overstretched states obtained by pulling from different ends. In 150 mM NaCl there is no hysteresis in the presence or absence of glyoxal for DNA pulled from the 3'3' ends. To probe the effect of glyoxal on the hysteresis of molecules pulled form the 3'3' ends, we measured the force vs. extension curves in 20 mM NaCl where we observed hysteresis in the absence of glyoxal. The presence of glyoxal had no effect on the hysteresis ($H_{Low}$ = 28 ± 16 pN μm) even for dsDNA pulled from the 3'3' ends again showing either that the glyoxal could not bind to this overstretched structure or that there is no significant amount of ssDNA available for binding.

    In the discussion above, we assumed that glyoxal had an effect on the hysteresis when the DNA was pulled from the 5'5' ends because it bound to small regions of ssDNA that were created by overstretching, but that either no ssDNA was created when the DNA was stretched from the 3'3' and 3'5' ends, or the binding of the glyoxal had no effect. In order to differentiate these two possibilities, we heated the 3'5' dsDNA molecules in 10 mM glyoxal in order to generate ssDNA to which the glyoxal can bind and determined whether the binding of glyoxal to ssDNA can affect the hysteresis for molecules pulled from the 3'5' ends. Figures 5 (a), (b), (c), (d), (e), and (f) show force vs. extension curves measured at room temperature for DNA stretched from the 3'5' ends in 10 mM glyoxal and 150 mM NaCl where the sample had previously been heated in the presence of glyoxal at 25, 35, 40, 60, 68, and 70ºC, respectively. The samples were cooled slowly to room temperature, where the force vs. extension curves were then taken. All of the samples had overstretching curves that overlapped if the force was cycled up and down several times. For these molecules that are pulled from the 3'5' ends, the molecules heated to 25 or 35ºC in glyoxal are hysteretic and similar to the overstretching curves taken in the absence of glyoxal; however, for molecules heated to 40ºC or above, there is no hysteresis. This suggests that some fraction of

ssDNA is opened at 40°C and that the glyoxal is able to bind and remain bound to these open regions. No significant ssDNA is visible until the sample is heated to temperatures above 65°C. Figure 4(d) shows the force vs. extension curve for a sample heated to 68°C where only about 50% of the base pairs undergo the overstretching transition, suggesting that the rest is ssDNA. For samples heated to 70°C, only a small fraction of the DNA remains double stranded, and the fraction undergoing the overstretching transition is between 10 and 20 % of the base pairs. Finally at 80°C in the presence of glyoxal the molecules show 100 % ssDNA formation as can be seen in Fig. 6. Notice that there is no hysteresis for any curve obtained for a sample heated above 40°C showing that the ssDNA opened by heating is held open by the glyoxal with no discernable force induced change in the ssDNA fraction. Thus, for molecules pulled from the 3'5' ends, a marked reduction in hysteresis occurs when ssDNA is created by melting the DNA in the presence of glyoxal. Figure 6 shows the ssDNA fraction as a function of the temperature to which the sample was previously heated in glyoxal. These results can be also fit to a transition equation similar to Eq. 5 to give $T_c$= 68°C and $\sigma$ =1.1°C, where the fraction of ssDNA at T = 0 ($frac_0$) and T = infinity ($frac_{infinity}$) are 0 and 1, respectively.

$$ssDNA_{fraction}[T] = frac_{\inf inity} + \left[ \frac{frac_0 - frac_{\inf inity}}{1 + \exp^{\frac{-T_c + T}{\sigma}}} \right] \qquad (6)$$

This curve clearly shows that there is not significant melting of the DNA in the glyoxal at 40°C, and no significant difference in the ssDNA fractions between 35°C and 40°C, despite the marked change in the hysteresis for the 3'5' construction with temperature where the curves for molecules previously heated in glyoxal at 35ºC were hysteretic, whereas those previously heated in glyoxal at 40ºC were not. This further suggests that there may be a conformational change in the dsDNA at temperatures between 35 and 40ºC, a result that would be consistent with the marked decrease in the unzipping force that occurs in the same temperature range [25].

**Summary**

We have considered the force vs. extension curves for dsDNA overstretched using different pulling techniques. We find that when the force increases with time, the force vs. extension curves are the same for all pulling techniques; however, we find that under some conditions, different pulling techniques produce consistent differences in the force vs. extension curves when the force decreases with time. These differences can be quantified in terms of the hysteresis in the force vs. extension curves. We demonstrate that when the maximum applied force is less than 1.2 times the overstretching force, less than 5% of the molecule is single stranded even for cases where the fraction of the molecule displaying hysteresis exceeded 10 %. Thus, the hysteresis is not due to the formation of ssDNA that fails to reanneal, rather it is due to the metastability of the overstretched conformation of dsDNA. In 150 mM NaCl, the force vs extension curves for molecules overstretched using the 3'3' ends show no hysteresis, whereas the curves for molecules overstretched using 5'5' ends or the 3'5'ends show a significant hysteresis. The measured hysteresis depends strongly on the ionic environment: all pulling techniques exhibit metastability in 20 mM NaCl, whereas no pulling method displays metastable states in 1 M NaCl. This difference is consistent with the suggestion that a metastable state will form if the Debye screening length exceeds the minimum separation between phosphates in the overstretched dsDNA configurations predicted by theory [13].

Additional differences between the overstretched conformations are observed when the molecules are overstretched in a buffer that contains glyoxal, which is a molecule that binds to ssDNA. For the 3'5' and 5'5' curves, the hysteresis in a 150 mM NaCl buffer can be reduced by binding glyoxal to ssDNA, whereas the glyoxal has no significant effect on the hysteresis when the DNA is pulled from the 3'3' ends. In the 5'5' case, the ssDNA to which the glyoxal binds can be created by overstretching from the 5'5' ends to a maximum force below 1.2 $F_{ov}$; however in the 3'5' case the presence of glyoxal in solution has no effect on the hysteresis. In contrast, if ssDNA is generated by heating the DNA in a buffer containing glyoxal to a temperature of at least 40ºC , then the force vs extension curves that are subsequently obtained by pulling from the 3'5' ends at room temperature show no hysteresis, though heating the DNA to 35 C in the same buffer has almost no effect on the hysteresis. Thus, the binding of the glyoxal to the ssDNA generated by heating to 40 C can remove hysteresis for a molecule subsequently overstretched from the 3'5' ends in a 150 mM NaCl buffer at room temperature, even though no significant ssDNA is detected in the heated molecules unless the temperature to which they are heated exceeds 60C. This result may be related to the dramatic decrease in the force required to unzip that occurs when the temperature of the DNA is increased from 35 C to 40C. The origin of these temperature dependent changes in the stability of DNA requires further investigation.

## Acknowledgements

We would like to acknowledge contributions by Julea Vlassakis, Jeremy Williams, Kristi Hatch, and Alyson Conover. This research was funded by grants: ONR DARPA N00014-01-1-0782; Materials Research Science and Engineering Center (MRSEC): NSF # DMR 0213805 and Army Research Office: grant W911NF-04-1-0170. N.K.'s research is supported by a grant from the NIH: RO1-GMS25326.

**LEGENDS**

FIG. 1. Stretching several dsDNA single molecules by pulling from different ends. Extension curves in different ionic strength conditions and 10 mM phosphate buffer pH 7.4 are shown from row 1 through 3: 1 M , 150 mM, and 20 mM NaCl, respectively. The columns from left to right show the results for each dsDNA construct that was pulled from the 3'3', 3'5', and 5'5' ends, respectively. In these curves the solid symbols and solid lines correspond to the curve where the force is increased whereas the hollow symbols and dashed lines are for the curves where the force was decreased.

Table 1: Summary of the hysteresis calculations for each pulling technique and each ionic strength condition. The hysteresis was calculated as the difference in the integral of the force vs. extension curve for increasing force and decreasing force for many different single molecules (between 5 and 10).

FIG. 2 Overstreching at forces above $F_{ov}$ (a). Force vs. extension curves for dsDNA pulled from the 3'5' ends in 150 mM NaCl and 10 mM phosphate buffer pH 7.4. In each cycle the final applied force is larger than in the previous one in order to achieve ssDNA formation represented by a change in slope for the cycle coming back and finally the purple curve overlaps the typical curve for ssDNA. (b) ssDNA fraction vs. force data corresponding to the shear transition that occurs during the increasing force portion of cycle 3 (squares) and fit to the data using Eq. 5 (dotted line) with $F_c$ =152 pN and $\sigma$ = 12 pN. (c) Data for the third cycle (blue circles on solid and dotted lines) shown in a) and WLC fit to the dsDNA below the overstretching transition and the same curve displaced to match the overstretched length above the overstretching transition. The solid purple line shows the theory that includes the fits to both the overstretching transition and the shear force transition whereas the red dotted line is the fit to ssDNA obtained from Eq. 4.

FIG 3 Force vs. extension curves for dsDNA pulled from the 3'3' ends (Klenow- biotin construct) (a) 150 mM NaCl and 10 mM phosphate buffer. (b) 0.5 M glyoxal, 150 mM NaCl, and 10 mM phosphate buffer (c) ssDNA fraction vs. maximum force using the data from a) (squares) and a fit to this data (dotted line) using Eq. 5 with $F_c$ = 130.7 pN $\sigma$ =11 pN, and $frac_{infinity}$ = 0.53.

FIG 4 Force vs. extension curves for three different single molecules stretched from the 5'5' ends. The solid lines and closed symbols represent the increasing force curve and the dashed lines and open symbols, the corresponding decreasing force curve. The blue and red curves were taken in 50 and 10 mM glyoxal in PBS buffer, respectively. The green curve is a control for 5'5' construct in the same buffer but in the absence of glyoxal.

FIG 5 Force vs. extension curves for dsDNA molecules pulled from the 3'5' ends previously heated in 10 mM glyoxal and finally overstretched at room temperature. Each sample was heated to a different maximum temperature: (a) 25, (b) 35, (c) 40, (d) 60, (e) 68, and (f) 70ºC.

FIG 6 ssDNA fraction as a function of the temperature to which the sample was previously heated in glyoxal (squares) and fit to the transition equation Eq. 6 (dotted line) with $T_c$ = 68°C, $\sigma$ =1.1°C, $frac_0$ = 0, and $frac_{infinity}$ = 1.

.

|       | $H_{Low}$ (pN μm) 20 mM | $H_{Low}$ (pN μm) 150 mM | $H_{Low}$ (pN μm) 1 M |
|-------|-------------------------|--------------------------|-----------------------|
| **3'3'** | 44 ± 36              | 0.3 ± 0.8                | 0.5 ± 0.53            |
| **5'5'** | 49 ± 30              | 13 ± 8                   | 1.1 ± 1.4             |
| **3'5'** | 25 ± 13              | 6 ± 4                    | 1.5 ± 1.7             |

Limouse et al. Table 1

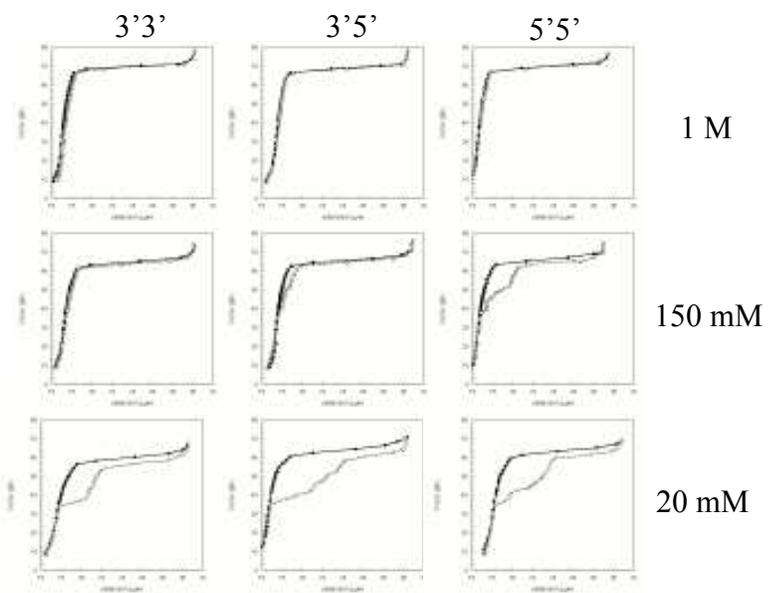

Limouse et al. Fig. 1

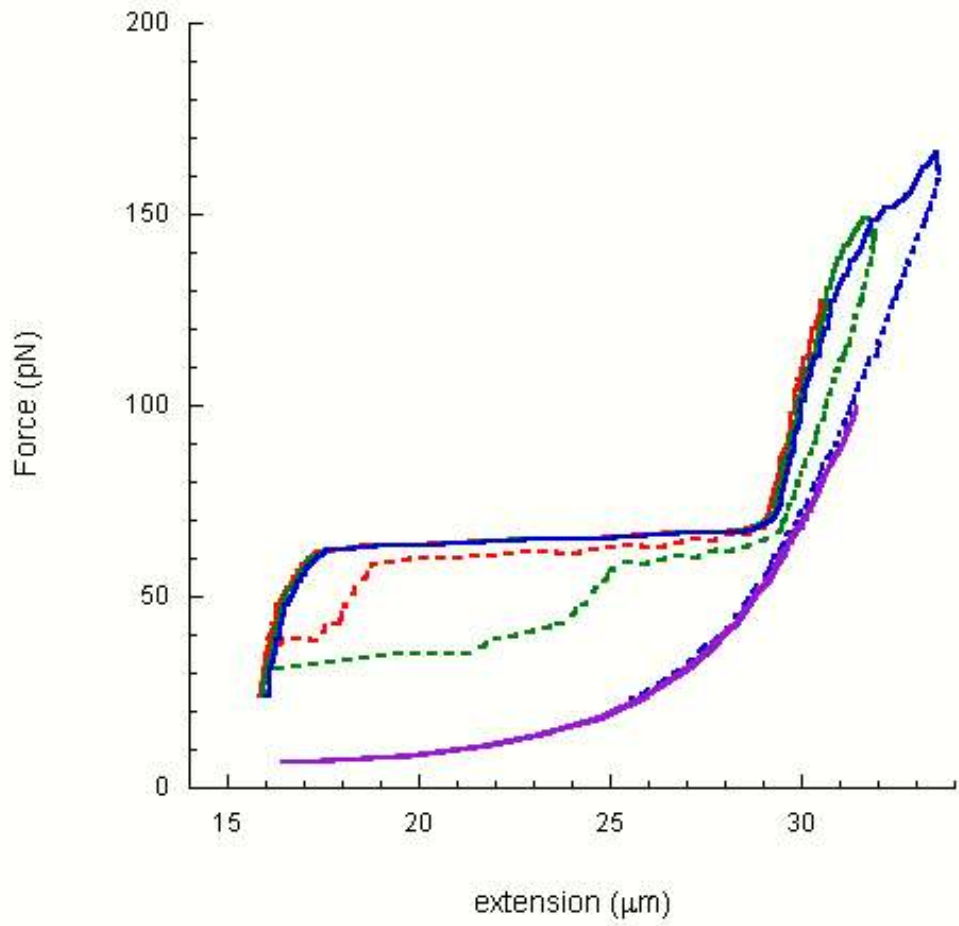

Limouse et al. Fig. 2(a)

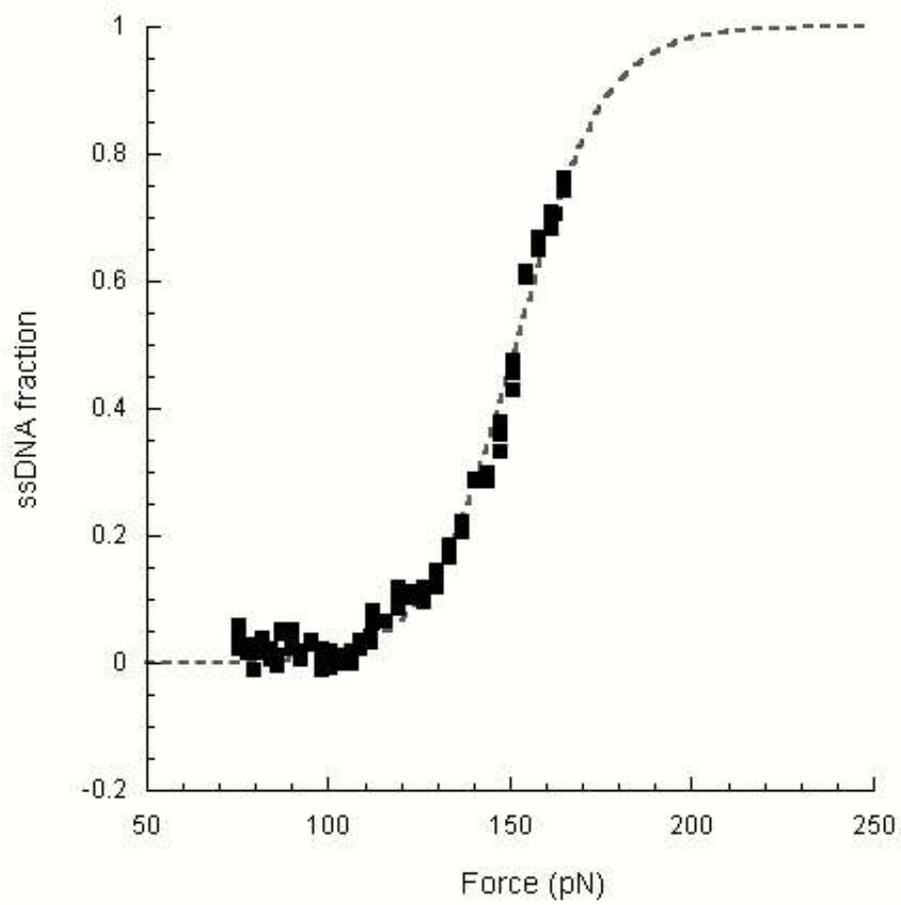

Limouse et al. Fig. 2(b)

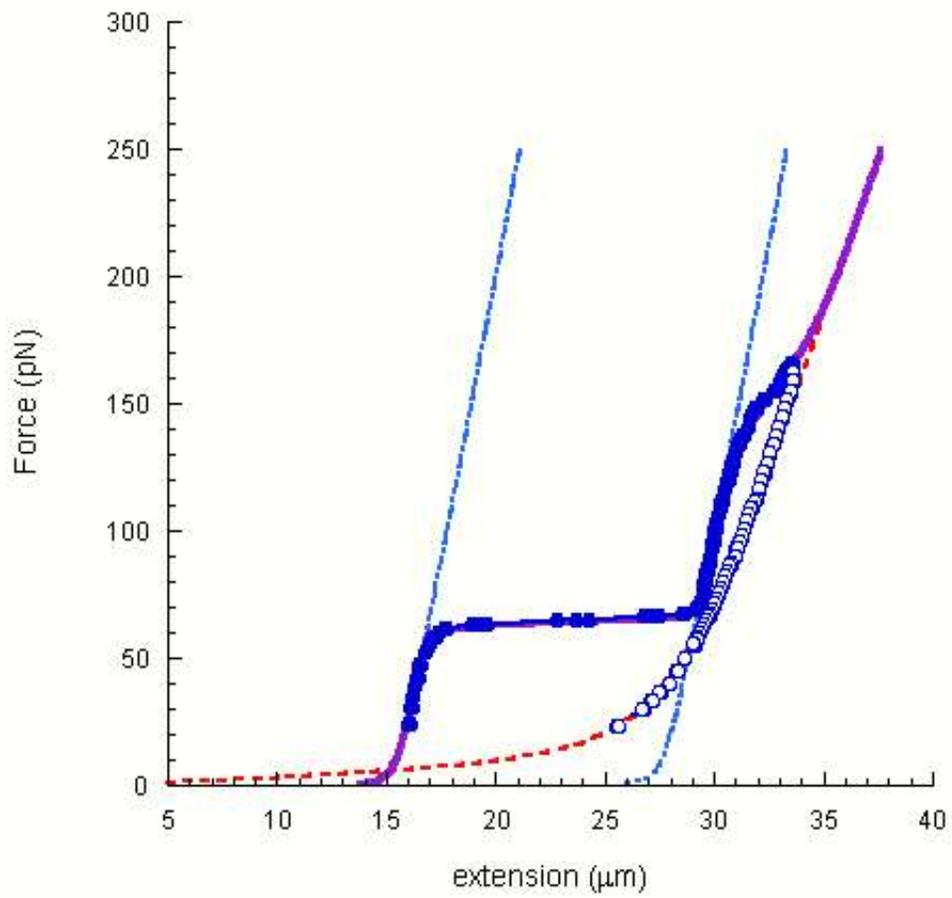

Limouse et al. Fig. 2(c)

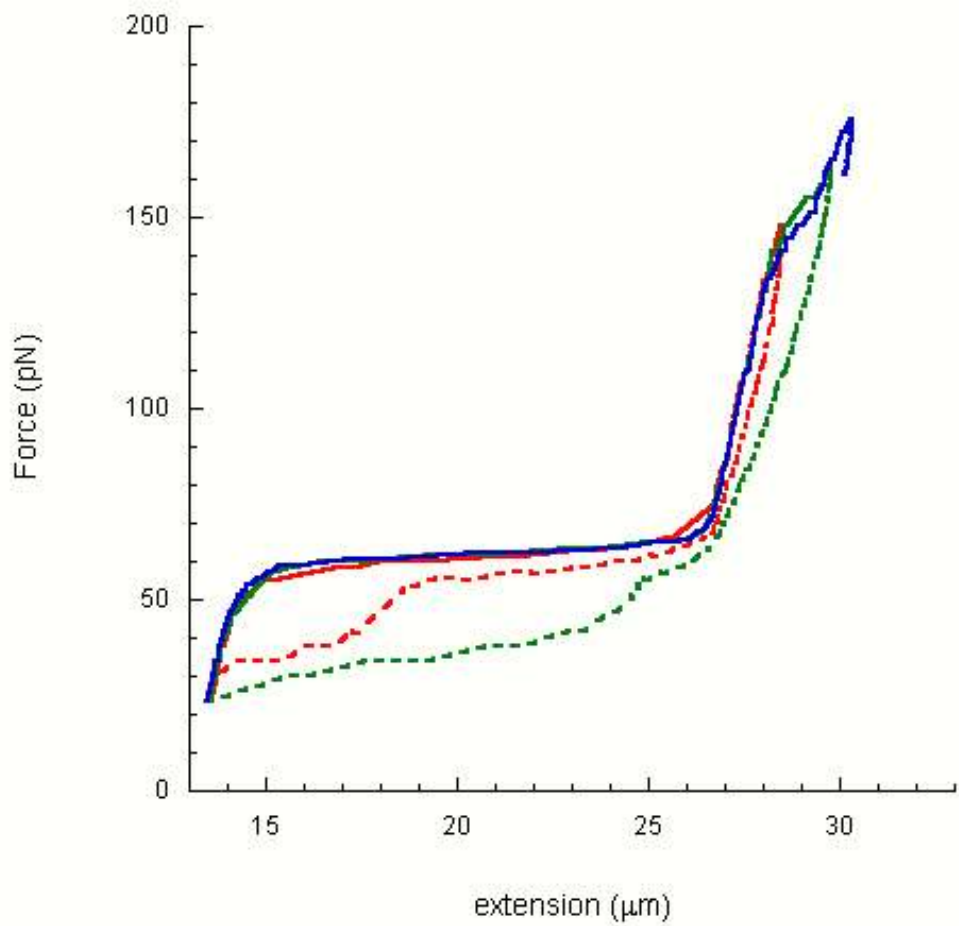

Limouse et al. Fig. 3(a)

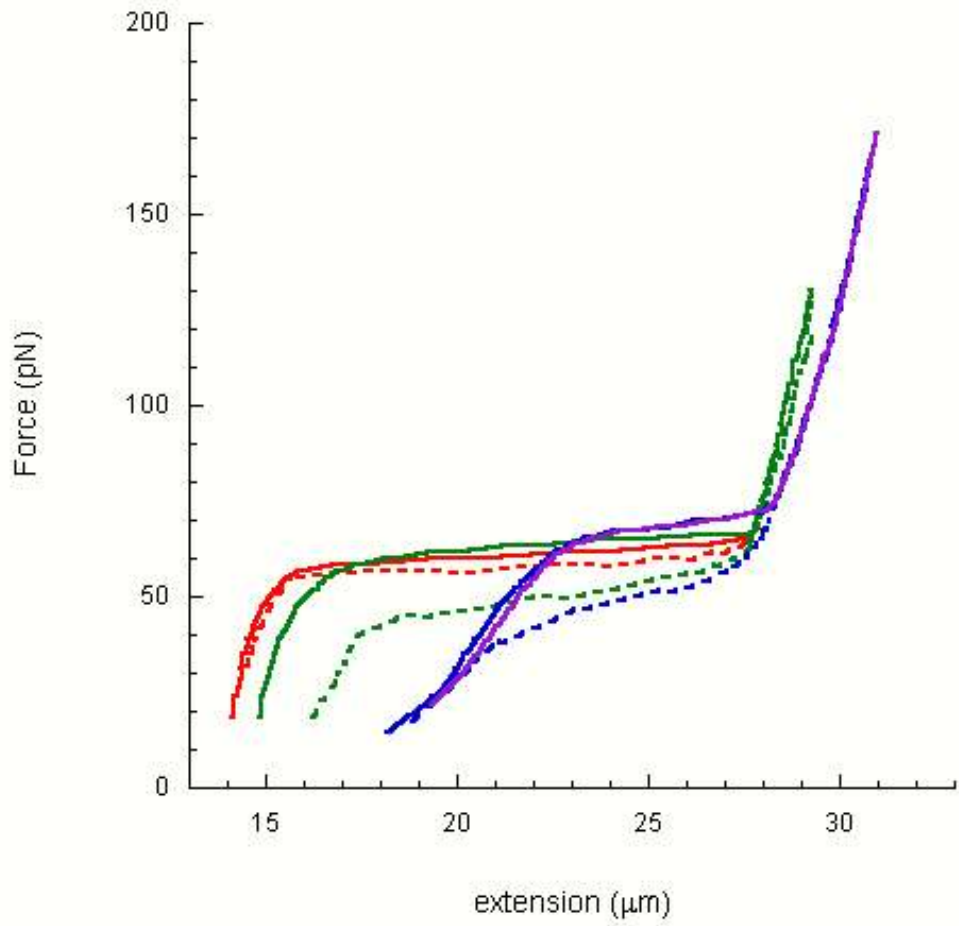

Limouse et al. Fig. 3(b)

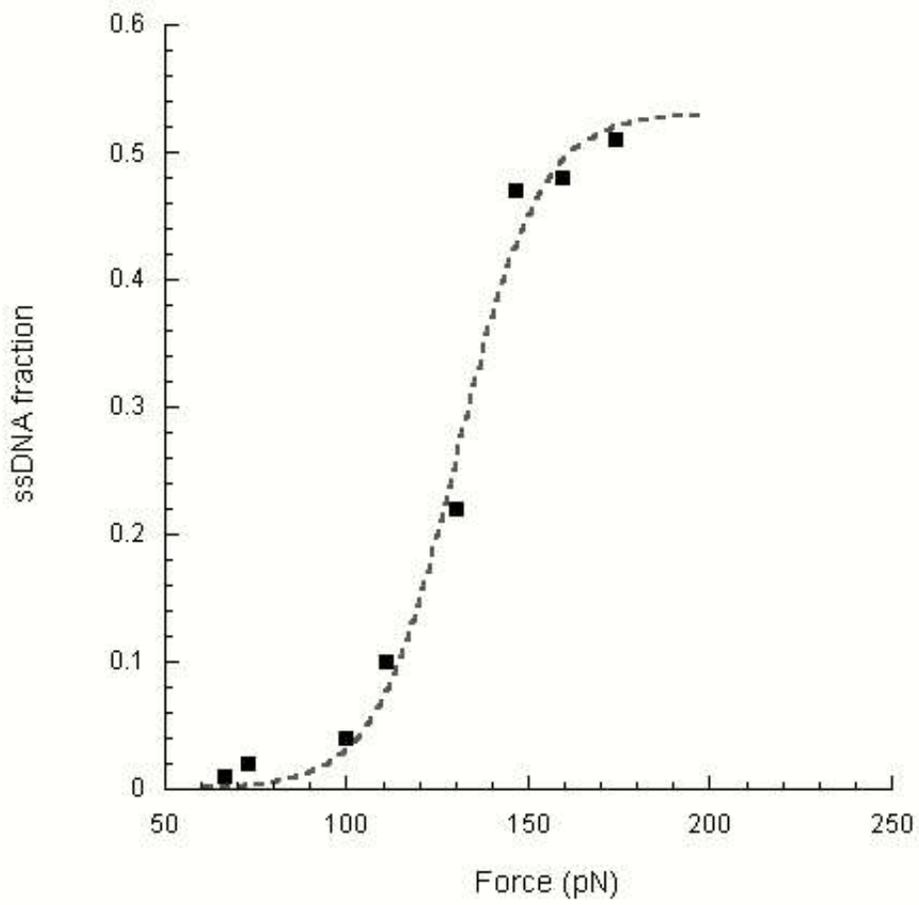

Limouse et al. Fig. 3(c)

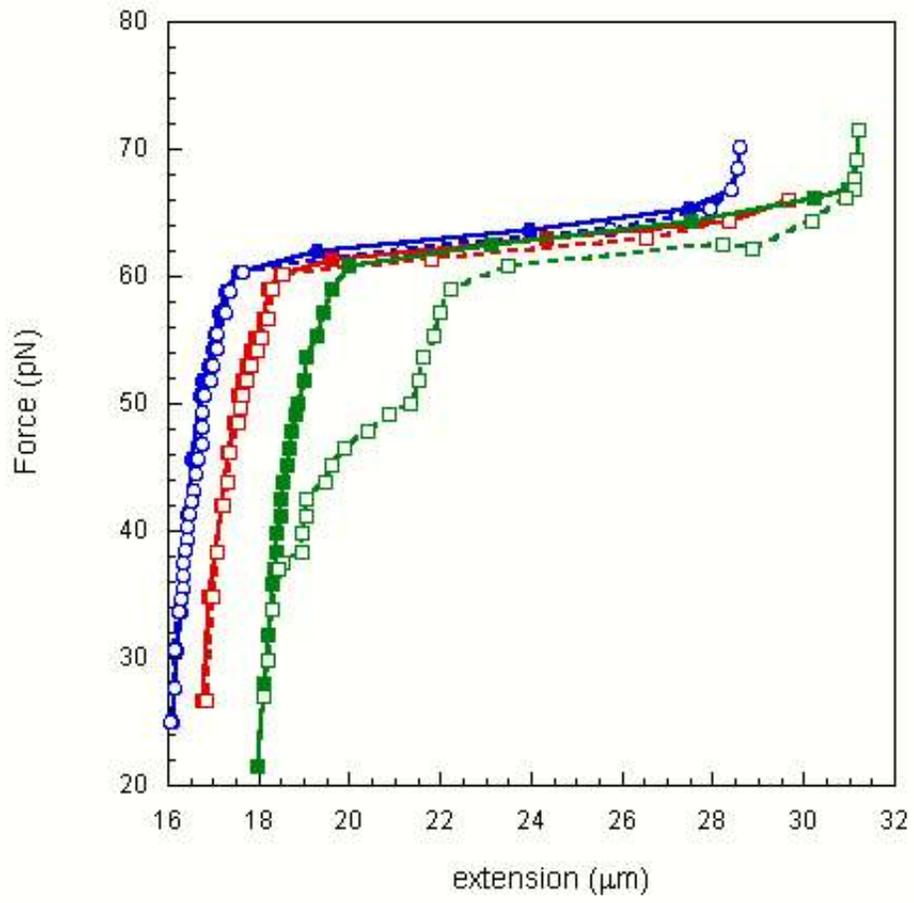

Limouse et al. Fig. 4

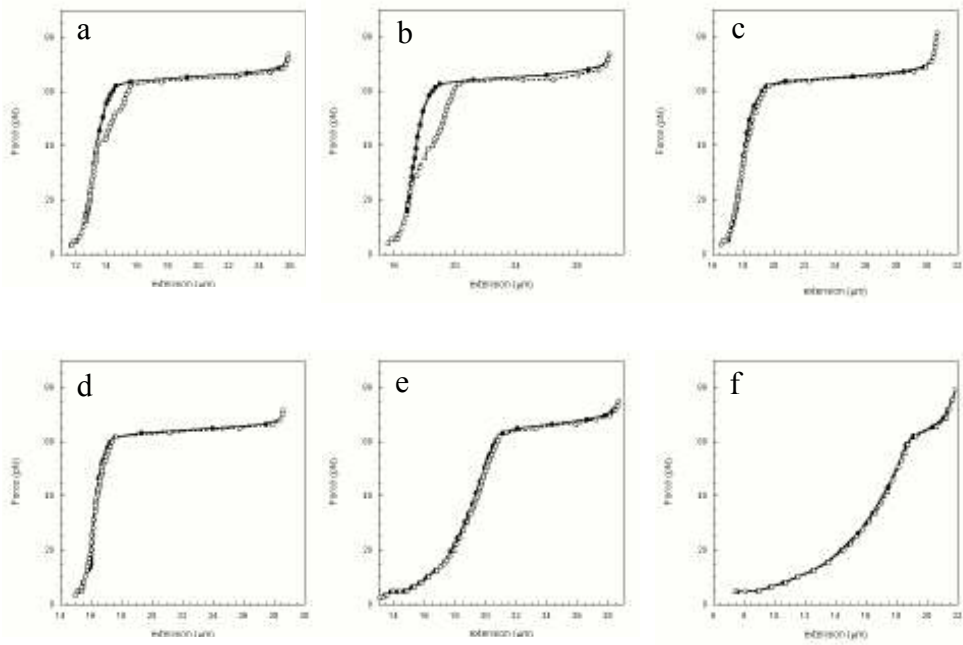

Limouse et al. Fig. 5

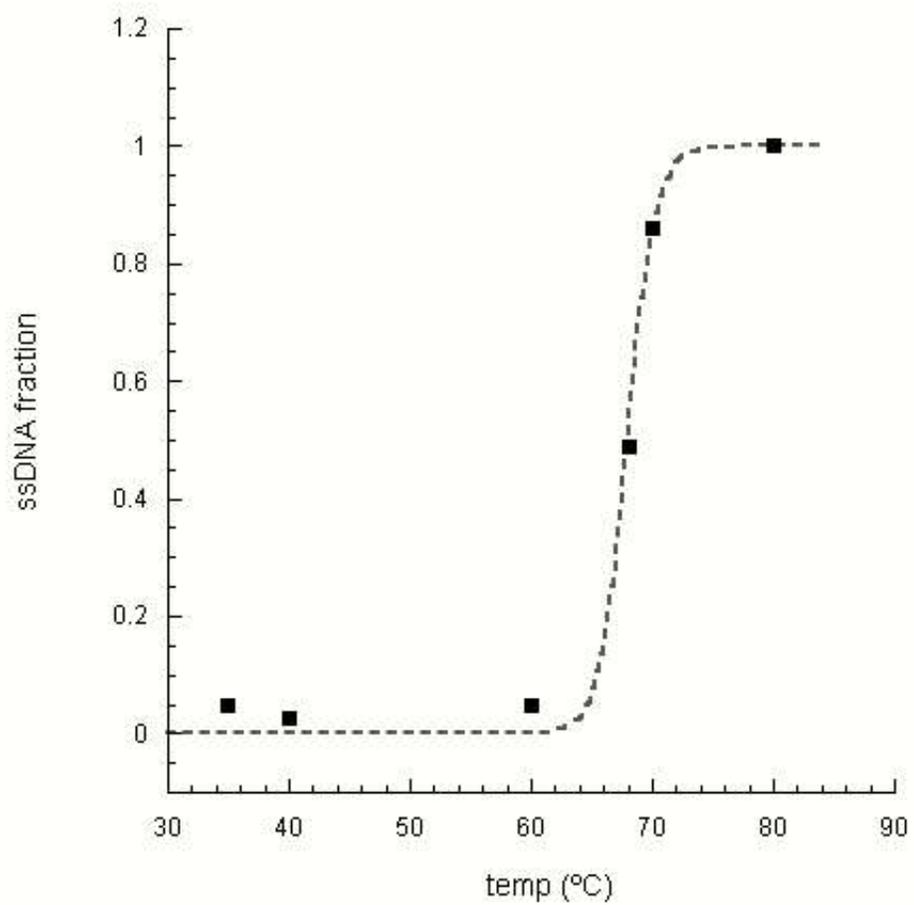

Limouse et al. Fig. 6